\documentclass[twocolumn,showpacs,prl,superscriptaddress]{revtex4-1}
\usepackage{graphicx}
\usepackage{color}\usepackage{enumerate}
\usepackage{amsmath}\usepackage{amssymb}\usepackage{stmaryrd}
\usepackage{multirow}
\usepackage{float}
\usepackage{longtable,booktabs}
\usepackage{subfigure}
\usepackage{dsfont}
\usepackage{amssymb}
\usepackage[pdftex,colorlinks=true,linkcolor=blue,citecolor=blue]{hyperref}

\begin{document}
\title{Topological Quadrupolar Semimetals}

\author{Mao Lin and Taylor L. Hughes}
\affiliation{Department of Physics and Institute for Condensed Matter Theory, University of Illinois at Urbana-Champaign, IL 61801, USA}
\begin{abstract}
In this work we predict several new types of topological semimetals that exhibit a bulk quadrupole moment. These semimetals are modeled with a 3D extension of the 2D quadrupole topological insulator. One type of semimetal has bulk nodes and gapped, topological surfaces. A second type, which we may call a higher order topological semimetal, has a gapped bulk, but harbors a Dirac semimetal with an even number of nodes on one or more surfaces. The final type has a gapped bulk, but harbors half of a Dirac semimetal on multiple surfaces. Each of these semimetals gives rise to mid-gap hinge states and hinge charge, as well as surface polarization, which are all consequences of a bulk quadrupole moment. We show how the bulk quadrupole moments of these systems can be calculated from the momentum-locations of bulk or surface nodes in the energy spectrum. Finally, we illustrate that in some cases it is useful to examine nodes in the Wannier bands, instead of the energy bands, to extract the bulk quadrupole moment.
\end{abstract}

%\pacs{}
\maketitle
The recent theoretical prediction of a new class of (higher-order) topological insulators with quantized quadrupole moments\cite{benalcazar2017A} has opened a new direction in the field of topological phases\cite{,peng2017quad,benalcazar2017B,song2017,neupert2017,langbehn2017,imhof2017}. The simplest quadrupole topological insulator is a 2D  system with an energy gap in the bulk and on the boundaries. This is unusual for topological insulators as they conventionally have characteristic gapless surface states\cite{hasan2010,bernevigbook}. However, the boundaries in the quadrupole insulator are not inert, and actually form lower-dimensional topological phases themselves. One manifestation of the surface topology is the existence of protected, mid-gap modes on the corners of the system where two edges intersect\cite{benalcazar2017A}. 

Extensions of this topological phase to three dimensional materials include new phases like topological octupole insulators\cite{benalcazar2017A,benalcazar2017B}, and topological insulators with chiral/helical modes on the surface hinges where two faces intersect\cite{benalcazar2017B,song2017,neupert2017,langbehn2017}. In this article we instead turn our attention to the prediction of new classes of topological \emph{semimetals} (TSMs) based on an extension of the quadrupole insulator to a layered 3D system. The new classes of TSMs include a bulk quadrupolar TSM with gapless bulk nodes, but without gapless surface modes, and several types of \emph{higher-order} TSMs (defined below) that are gapped in the bulk, but harbor surface TSMs. Each of these 3D TSMs has a quadrupole moment that can be determined by the geometry of the bulk or surface point-node band-crossings in the system; this is analogous to the electromagnetic response properties of 3D Weyl semimetals\cite{wan2011,yang2011,burkov2011weyl,zyuzin2012,vazifeh2013,ramamurthy2015} and 2D/3D Dirac semimetals\cite{ramamurthy2015} which can be determined by the location of the Weyl/Dirac nodes in energy/momentum space.

This article is organized as follows: (i) we will review some basic aspects of the simple quadrupole model presented in Ref. \onlinecite{benalcazar2017A} including the phase diagram, (ii) we stack the quadrupole model into a layered geometry and introduce coupling between the layers in order to generate 3D TSMs, (iii)  finally we will illustrate the phenomenology associated with several new classes of TSMs, and show how the quadrupole moment of each system is tied to the momentum-space locations of the gapless nodal points. 

\begin{figure}
\centering
\includegraphics[width=\columnwidth]{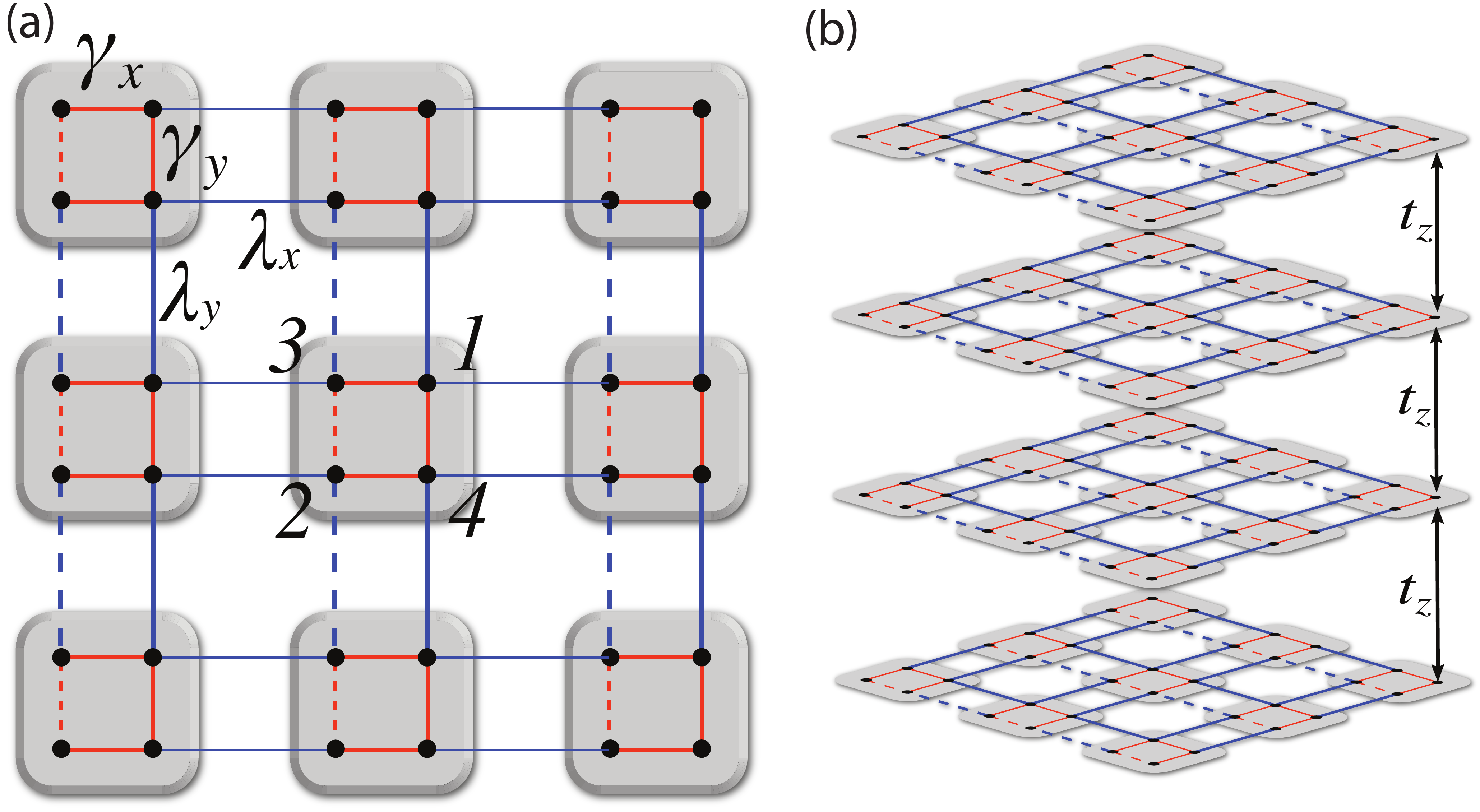}
\caption{(a) Tightbinding representation of the 2D topological quadrupole insulator in Eq. \ref{eq:C4_quadrupole}. Each black dot represents a single spinless electronic orbital.  Each solid line represents a tunneling term. Each dotted line corresponds to a hopping with the same strength as a corresponding solid line, but with a relative phase of $-1$ which is a gauge choice that inserts a $\pi$-flux in each plaquette (including within the unit cell). The ordered basis for the $\Gamma$-matrices is shown on the central plaquette. (b) Stacking the 2D quadrupole insulator into a 3D system and coupling the layers to generate the TSM model Eq. \ref{eq:stack_mirror_quadrupole}.}
\label{fig:1}
\end{figure}

Let us begin with a model for a 2D topological quadrupole insulator\cite{benalcazar2017A}. A tightbinding representation of the Hamiltonian is illustrated in Fig. \ref{fig:1}a with four spinless orbitals per unit cell, and includes inter- and intra-cell nearest neighbor hopping. The Bloch Hamiltonian is
\begin{eqnarray}\begin{aligned}
\label{eq:C4_quadrupole}
H({\bf k})&=(\gamma_x+\lambda_x\cos k_x)\Gamma_4+\lambda_x\sin k_x\Gamma_3\\
&\quad+(\gamma_y+\lambda_y\cos k_y)\Gamma_2+\lambda_y\sin k_y\Gamma_1,
\end{aligned}\end{eqnarray}
where $\Gamma_0=\tau_3\otimes\mathbb{I}$, $\Gamma_k=-\tau_2\otimes\sigma_k,$ $\Gamma_4=\tau_1\otimes\mathbb{I},$ $\mathbb{I}$ is the $2\times 2$ identity matrix, and $\tau_a,$ $\sigma_a$ are Pauli matrices with a basis specified in Fig. \ref{fig:1}a. $\gamma_i$ and $\lambda_i$ are intra- and inter-cell tunneling strengths. There is $\pi$-flux per plaquette, and we have made a gauge choice for the relative phases of the hopping terms as shown in Fig. \ref{fig:1}a. For all values of $\gamma_i$ and $\lambda_i$ the model has $x$ and $y$ mirror symmetries with representation matrices $\hat{m}_x=\tau_1\otimes \sigma_3,$ and $\hat{m}_y=\tau_1\otimes\sigma_1.$ Due to the $\pi$-flux, these mirror operators \emph{anticommute}. If $\vert\gamma_x\vert=\vert\gamma_y\vert,$ and $\vert\lambda_x\vert=\vert\lambda_y\vert$ then the model has $C_4$ rotation symmetry with matrix representation
\begin{eqnarray}\begin{aligned}
\hat{r}_4=\begin{pmatrix}
0 & \mathbb{I} \\
-i\sigma_2 & 0
\end{pmatrix},
\end{aligned}\end{eqnarray}\noindent where we note that $\hat{r}_4^4=-1$ due to the $\pi$-flux.  

\begin{figure}
\centering
\includegraphics[width=\columnwidth]{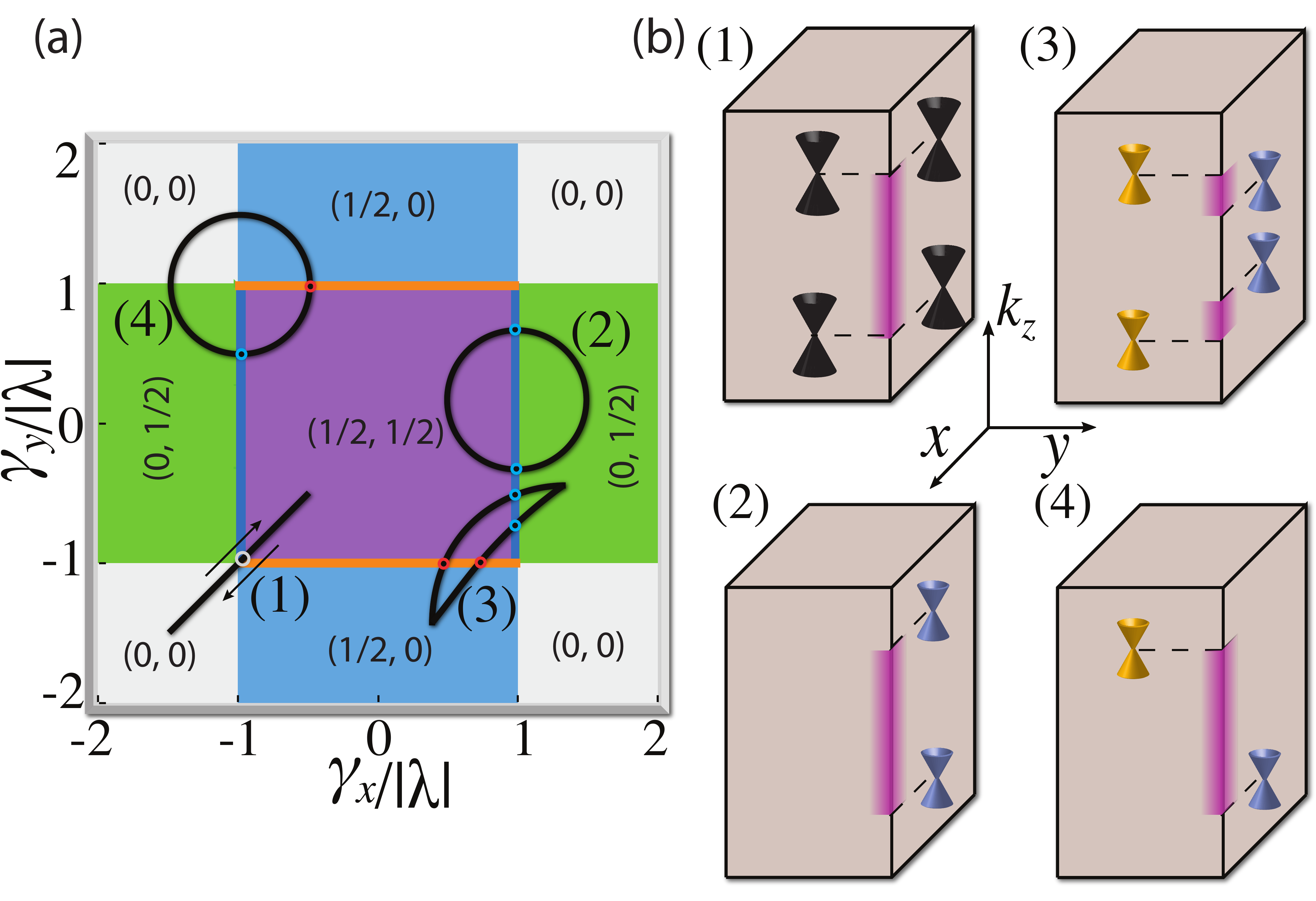}
\caption{(a) Phase diagram of the 2D quadrupole insulator model Eq. \ref{eq:C4_quadrupole} as a function of $(\gamma_x/\vert\lambda\vert, \gamma_y/\vert\lambda\vert)$. The ordered pair in each colored region represents the $\mathbb{Z}_2\times\mathbb{Z}_2$ topological class specified by the quantized Berry phases $(p_x, p_y)$ of the Wannier bands.  Each of the four paths represents a different type of parameterization of the TSM model in Eq. \ref{eq:stack_mirror_quadrupole}. (b) To each of the four paths there is an associated set of bulk (1) or surface (2), (3), (4) gapless nodes. Note that the horizontal directions represent spatial dimensions, while the vertical direction is $k_z$ momentum. The locations of hinge modes corresponding to each TSM configuration are highlighted on one hinge of the sample. The color of the node corresponds to the type of phase transition from which it is generated: black nodes are bulk transitions while the blue and orange nodes correspond to transitions in the Wannier bands when passing from the $(1/2, 1/2)$ class to the $(1/2, 0)$ [orange] or $(0, 1/2)$ [blue] classes. Note that the actual path 3 parameterization used in our numerics is a degenerate line similar to path 1. We show a more open path here for illustration. }
\label{fig:2}
\end{figure}

For our discussion of TSM phases it is important to understand the phase diagram of this model, shown in Fig. \ref{fig:2}a, as a function of the $\gamma_i$ and $\lambda_i$. To simplify the discussion let us fix $\lambda_x=\lambda_y=\lambda$ from now on. Throughout the entire phase diagram, the system has the pair of $x, y$ mirror symmetries, and on the diagonal and anti-diagonal it has $C_4$ symmetry. The model exhibits a $\mathbb{Z}_2\times \mathbb{Z}_2$ set of topological classes specified by the polarizations (Berry-Zak phases) $(p_x, p_y)$ of the hybrid Wannier bands $\nu_{y}(k_x)$ and $\nu_{x}(k_y)$ respectively\cite{benalcazar2017A,benalcazar2017B}. The $\nu_{i}(k_j)$ are Wannier center spectra resolved along the $i$-th direction; they take values between $0$ and $1,$ depend on momenta transverse to $\hat{i}$, and are calculated from the spectra of a Wilson loop in momentum-space oriented parallel to $\hat{i}$ (for extensive discussions see Refs. \onlinecite{benalcazar2017A,benalcazar2017B}). The $(p_x, p_y)$ Berry phases are quantized by the mirror symmetries, and take values of either $0$ or $1/2$ in units of $2\pi.$ The interior (purple) square of the phase diagram represents the topological quadrupole phase having $(p_x, p_y)=(1/2,1/2).$ This is the only region of the phase diagram with a non-vanishing quadrupole moment $q_{xy}$; note that $q_{xy}$ is quantized by either the mirror or $C_4$ symmetries and hence only takes the values $0$ or $1/2$ throughout the entire phase diagram. 

 If $C_4$ symmetry is enforced, then in this phase diagram there is only one type of possible phase transition: from the quadrupole $(1/2, 1/2)$ phase to the fully trivial phase $(0,0).$ The transition occurs when $\vert \gamma\vert=\vert\lambda\vert,$ at which there is a four-band Dirac node in the 2D bulk energy bands at one of the time-reversal invariant momenta; the momentum location depends through which corner of the topological quadrupole region the parameters are tuned\cite{benalcazar2017B}. 

When $C_4$ symmetry is relaxed, two other classes of transitions are available. When one passes from the interior square horizontally (vertically) to one of the green (blue) regions by increasing $\vert\gamma_x/\lambda\vert$ ($\vert\gamma_y/\lambda\vert$) while keeping $\vert \gamma_y/\lambda\vert <1$ ( $\vert \gamma_x/\lambda\vert <1$), then there will be a gap-closing transition in the Wannier bands, but not the bulk energy bands\cite{benalcazar2017A,benalcazar2017B}. Interestingly, when leaving the $(1/2, 1/2)$ phase the Wannier transition is accompanied by a gap-closing in the \emph{edge} energy spectrum when the system has open boundary conditions\cite{benalcazar2017A,benalcazar2017B}. Specifically, when $\vert\gamma_x/\lambda\vert=1$ ($\vert\gamma_y/\lambda\vert=1$) and $\vert\gamma_y/\lambda\vert<1$ ($\vert\gamma_x/\lambda\vert<1$), there will be a gap-closing in the $\nu_y (k_x)$ ($\nu_x(k_y)$) Wannier bands at a value of $\nu_{y}=1/2$ ($\nu_{x}=1/2$) that results in an energy-gap closing transition for edges parallel to $\hat{x}$ ($\hat{y}$). These transitions take the system from the $(1/2, 1/2)$ class to the $(0, 1/2)$ ($(1/2, 0)$) class. In contrast, transitions from the $(1/2, 0)$  ($(0, 1/2)$) class to the $(0, 0)$ class have a gap-closing in the hybrid-Wannier bands at a value of  $\nu_{y}=0$ ($\nu_{x}=0$), but there is not a generic gap-closing transition in the bulk or edge energy spectrum. We also mention that despite having non-trivial Wannier-band topology, the $(1/2, 0)$ and $(0, 1/2)$ classes do not have any physical manifestation of a quadrupole moment. 

The important features to take away from this discussion are: (i) only the $(1/2, 1/2)$ class has a quadrupole moment, hence it is the only class with non-vanishing boundary polarizations and corner charges,  (ii) transitions into and out of the $(1/2, 1/2)$ phase are accompanied by an energy gap closing in the bulk if $C_4$ is preserved, or an energy-gap closing on the edge, and a gap closing in the Wannier bands, if only the mirror symmetries are preserved, and (iii) transitions into and out of the $(0, 0)$ phase from $(1/2, 0)$ or $(0, 1/2)$ are accompanied by gap-closing transitions in the Wannier bands, but not the bulk or edge energy bands.

Now let us stack the 2D quadrupole model and couple the layers to generate a Bloch Hamiltonian (see Fig. \ref{fig:1}b):
\begin{eqnarray}\begin{aligned}
\label{eq:stack_mirror_quadrupole}
H({\bf k})&=(\gamma_x+\chi_x(k_z)+\lambda\cos k_x)\Gamma_4+\lambda\sin k_x\Gamma_3\\
&\quad+(\gamma_y+\chi_y(k_z)+\lambda\cos k_y)\Gamma_2+\lambda\sin k_y\Gamma_1,
\end{aligned}\end{eqnarray}\noindent where the $\chi_j (k_z)$ are periodic functions on the $k_z$ Brillouin zone (BZ) determined by the choice of inter-layer tunneling terms.  To understand why this model can generate a TSM, consider the quantities $\gamma_i(k_z)\equiv \gamma_i+\chi_{i}(k_z),$ which represent maps from the $k_z$ BZ to closed paths in the 2D phase diagram in Fig. \ref{fig:2}a with base point $(\gamma_x/\vert\lambda\vert, \gamma_y/\vert \lambda\vert).$ In Fig. \ref{fig:2}a we have illustrated four different types of paths, each of which has some portion of the path within the topological quadrupole phase, and each having a discrete set of $k_z$ at which the path is at a transition point between different topological classes. This is precisely what is needed for a TSM, and how we define a TSM, i.e., there are transition points, as a function of momentum, between different topological classes (as a comparison, recall that for a Weyl semimetal the Weyl nodes represent transition points between momentum slices with different Chern number). Furthermore, for each value of $k_z$ the system has mirror symmetries, and thus the Bloch Hamiltonian at a fixed $k_z$ has a quantized quadrupole moment, if it is not at a transition point. Hence, the bulk quadrupole moment of each path can be straightforwardly calculated by summing up all of the values of $k_z$ that are mapped, via the $\gamma_{i}(k_z),$ into the topological quadrupole region of the phase diagram. We will see below how, for each path, this calculation can be recast in terms of the momentum space locations of the transition points, analogous to, say, the calculation of the anomalous Hall coefficient in Weyl semimetals which is proportional to the momentum separation of the Weyl nodes\cite{wan2011,yang2011,burkov2011weyl,zyuzin2012,vazifeh2013}.  Let us move on to to describe the phenomenology of each path in turn. 

\begin{figure}
\centering
\includegraphics[width=\columnwidth]{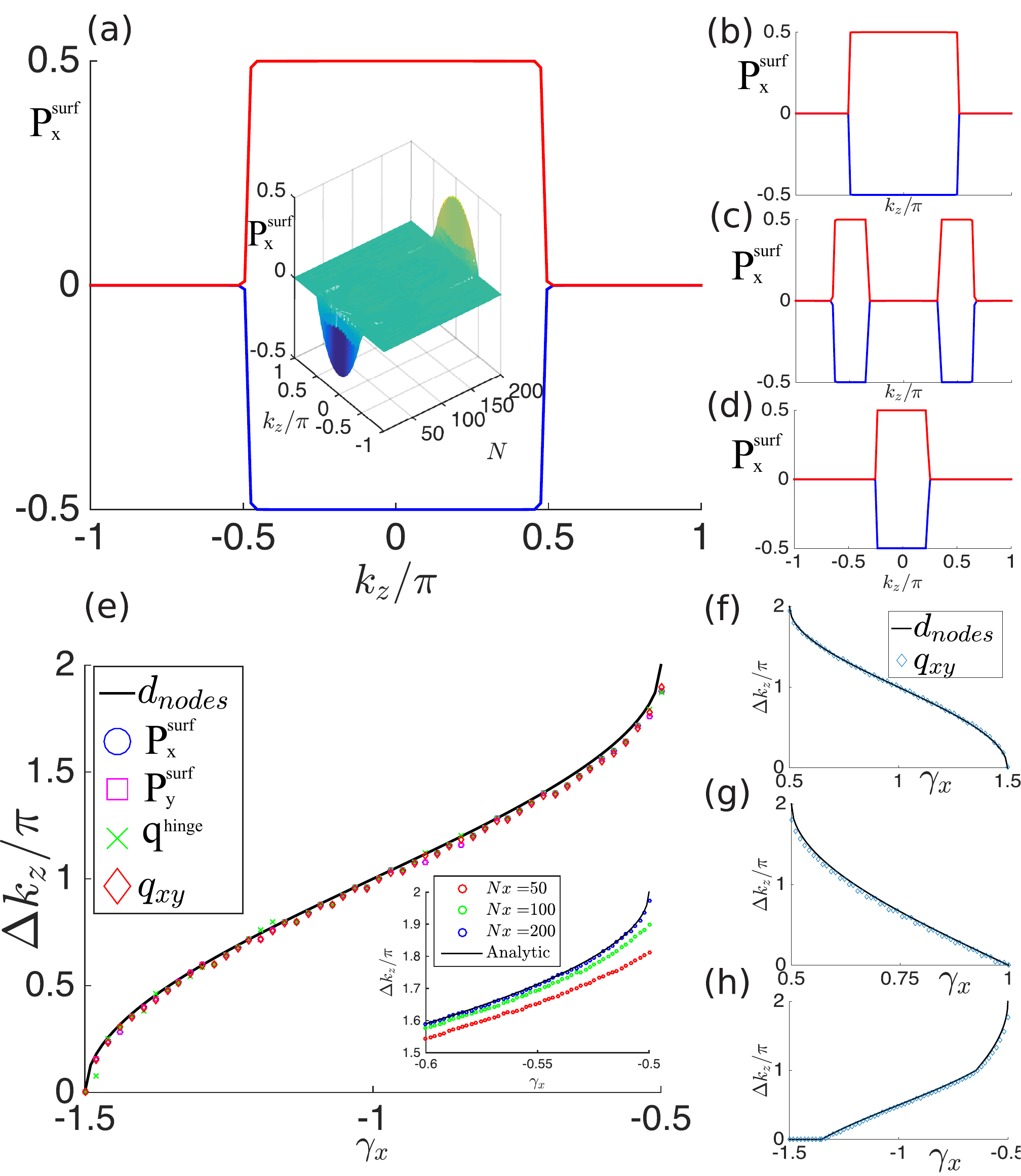}
\caption{Edge polarization vs. $k_z$ for paths (a) 1 (b) 2 (c) 3 (d) 4 for a system size of 200 in the (real-space) x-direction. Red and blue lines indicate polarization on opposite edges. Inset in (a) shows polarization resolved vs. $k_z$ and the spatial direction with open boundary. (e) Comparison of nodal separation, surface polarizations (system size 100 in the (real-space) x-direction), hinge charge per unit length, and quadrupole moment for path 1 as the base point is varied with $\gamma_x=\gamma_y.$ Setting the electric charge $e=1,$ all quantities have units of inverse length. Inset shows scaling of the surface polarization at the upper end of the range of $\gamma_x$ as system size is increased. The results converge systematically to the analytic result.  (f), (g), (h) show comparison of nodal separation and quadrupole moment for paths 2, 3, 4 respectively as the base points are varied with (f) $\gamma_y=1/4,$ (g) $\gamma_x=-\gamma_y,$ (h) $\gamma_x=-\gamma_y.$   }
\label{fig:3}
\end{figure}

In order to generate path 1 we need a $C_4$ invariant parameterization such as $\gamma_{i}(k_z)=-1+1/2\cos(k_z),$ for $i=x, y,$ and  $\lambda=1.$  This path passes through the $C_4$ invariant phase transition point when $\gamma_{i}(k_z)=-1$, i.e., when $k_z=\pm \pi/2.$ The two bulk nodes are represented by the black cones in Fig. \ref{fig:2}b. The nodes represent a transition, as a function of $k_z,$ between regions of the $k_z$ BZ in a topological quadrupole phase, and regions in the trivial phase. We can expand around one of the nodal points, say ${\bf{k}}=(0,0,\pi/2),$ to find the continuum Hamiltonian $H_{node}=\delta k_x\Gamma_3+\delta k_y\Gamma_1+(1/2)\delta k_z(\Gamma_2+\Gamma_4).$ This is a gapless Dirac Hamiltonian with a four-fold degenerate Dirac point when all $\delta k_i=0.$   Adding terms proportional to $\Gamma_0$ or $(\Gamma_2-\Gamma_4)$ will open a gap, but these terms are forbidden by a combination of mirror and $C_4$ symmetries. Hence, this TSM phase is protected by both mirror and $C_4.$ If one, for example, preserves $C_4$ but allows for the breaking of mirror symmetry, then one possibility is that the four-fold Dirac point could be split into two Weyl points (by adding $i m_{13}\Gamma_1\Gamma_3$), which would then produce a coexistence between a $k_z$ region with quadrupole topology and a $k_z$ region with vanishing quadrupole, but non-zero Chern number. We will leave a discussion of this hybrid Weyl-quadrupole semimetal to future work. 

The bulk quadrupole TSM generated by path 1 is similar to the usual Dirac and Weyl TSMs in that there are point nodes in the bulk spectrum, and one can interpret the TSM phase as an intermediate gapless phase between a weak topological insulator (built from a stack of lower dimensional topological insulators) and a trivial insulator\cite{yang2011,burkov2011weyl,ramamurthy2015}. Unlike the Dirac/Weyl system, however, this TSM does not have gapless Fermi-arc surface states. Instead, the surfaces are generically gapped unless they intersect another surface at a hinge. This is a consequence of the gapped, but topological nature of the quadrupole insulator edge states which have been stacked to form the surface states of the TSM. From our discussion above, we expect this system to have a bulk quadrupole moment $q_{xy}=\frac{e}{2\pi}b_z$ where $2b_z$ is the momentum space separation between the bulk nodes (the factor of 2 is chosen to match with the previous literature on TSMs). This momentum difference precisely accounts for the portion of path 1 in the topological quadrupole region of the phase diagram. As a result, this system will exhibit a surface polarization tangent to surfaces that are normal to the $x$- or $y$-directions, and hinge charges/mid-gap bound states on hinges where the polarized surfaces intersect. We confirm these results in numerical calculations in Fig. \ref{fig:3}a,e where we show the surface polarization resolved over the $k_z$ BZ, and show it is non-vanishing and quantized to $1/2$ in the region between the bulk nodes. We also shift path 1 by tuning $\gamma_x\in [-1.5, -0.5],$ while constraining $\gamma_y=\gamma_x$ and calculate the surface polarizations\cite{vanderbilt2015,benalcazar2017B}, hinge charge, and quadrupole moment $q_{xy}$ (via the nested Wilson loop method\cite{benalcazar2017A}) and show that they all match with the nodal separation formula above for the entire range of the $\gamma_i.$

\begin{figure}
\centering
\includegraphics[width=\columnwidth]{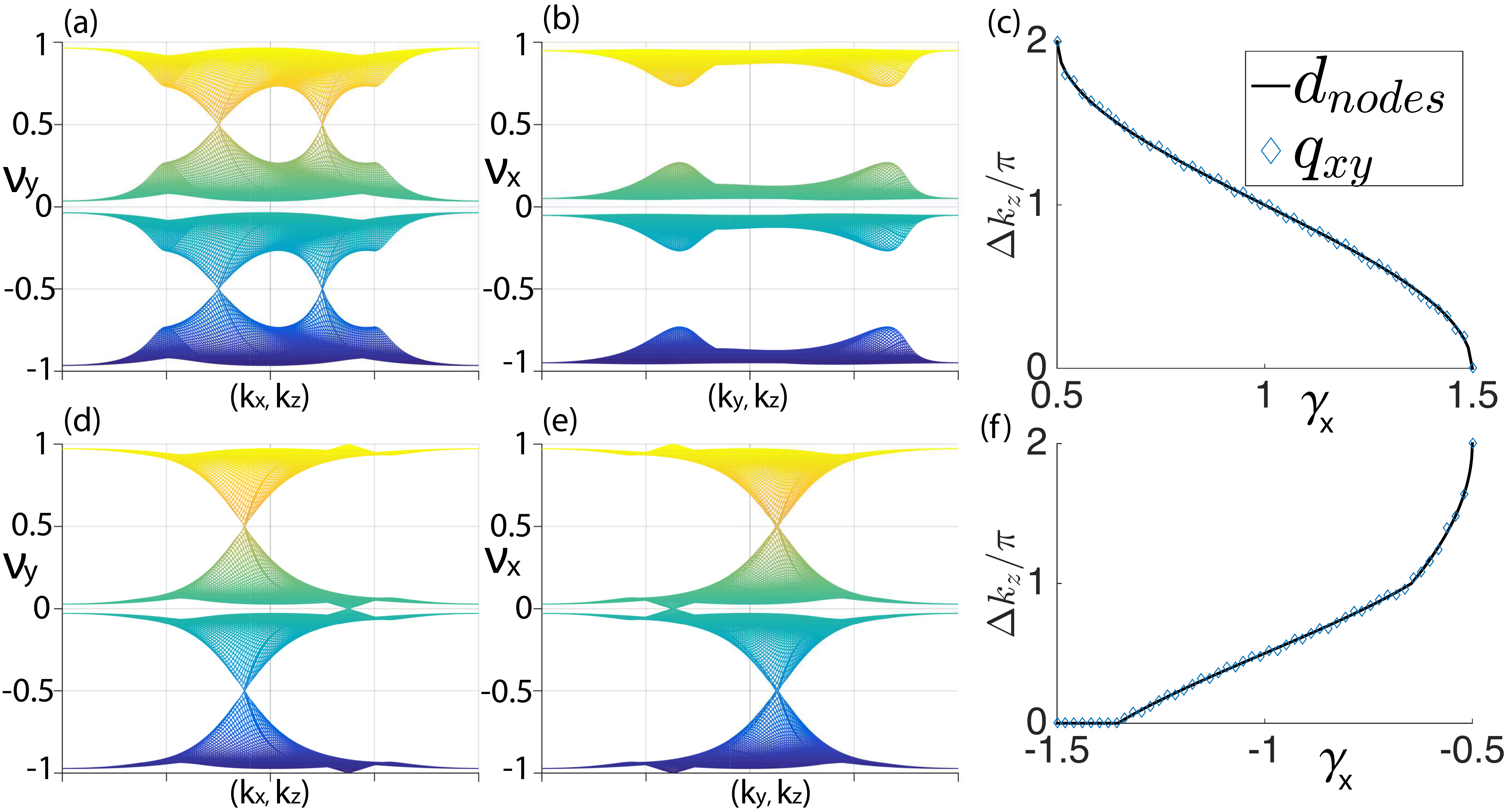}
\caption{Wannier bands in $x$ and $y$ directions calculate for path 2 (a),(b) and path 4 (d), (e) plotted vs. the momenta in the transverse BZ. Note that we have used a repeated zone scheme in the Wannier value direction for clarity. Path 2 has Wannier nodes at $\nu_{y}=1/2$, but no nodes in $\nu_x.$ Path 4 has nodes at $\nu_x=0, 1/2$ and at $\nu_y=0, 1/2.$ Nodes at a Wannier value of $1/2$ have corresponding nodes in the surface energy spectrum\cite{benalcazar2017B}. In (c), (f) we confirm that the difference in $k_z$ momenta of the nodes at a Wannier value of $1/2$ correctly reproduce the bulk quadrupole moment for (c) path 2 and (f) path 4 as the base points of these paths are varied identically to Fig. \ref{fig:3}f,h.}
\label{fig:4}
\end{figure}

While path 1 has a similar bulk-node structure to conventional TSMs, paths 2, 3, and 4 represent a completely new type of TSM, which we might call a higher-order TSM. All three of these paths are formed in regions of the phase diagram where only the mirror symmetries are generically preserved. Let us treat paths 2 and 3 first. We can parameterize path 2 via $\gamma_{x}=1+1/2\cos k_z, \gamma_y=1/4+1/2\sin k_z,$ and path 3 via $\gamma_{x}=3/4+1/2\cos k_z, \gamma_y=-3/4+1/2\cos k_z.$ Neither path hits a transition point where the bulk band gap closes, however there are two  and four points, for path 2 and 3 respectively, where the Wannier bands have a gap closing when leaving the $(1/2, 1/2)$ class (see Fig. \ref{fig:4}a,b for path 2). Consequently, for path 2, when the system has open boundaries there will be two values of $k_z$ at which the surface energy spectrum has a gap closing for surfaces normal to $\hat{y}.$ Alternatively we could have oriented path 2 so it hit one of the orange phase boundaries in Fig. \ref{fig:2}a, and subsequently we would find gapless nodes on surfaces normal to $\hat{x}.$ Path 3 is similar to two copies of path 2, one copy with each orientation, and it will have gapless points on surfaces normal to $\hat{x}$ and surfaces normal to $\hat{y}$ since it intersects both types of Wannier transition points as $k_z$ traverses the BZ. Hence, these systems are gapped in the bulk, but have TSMs on their surfaces.  Indeed these systems have surface Dirac semimetals with an even number of nodes (possibly zero) on each surface normal to $\hat{x}$ and/or $\hat{y}.$ 
The gapless nodes are protected by mirror symmetries and lie on mirror-invariant lines in the surface BZ. As an example, take path 2 and extract the low-energy surface Hamiltonian on the surface normal to $\hat{x}$\cite{konig2008}. The resulting continuum Hamiltonian, when expanded around a surface Dirac node, reads $H_\text{surf-node}(k_y,k_z)=v k_z\sigma_1-\lambda k_y\sigma_2,$ where $v$ depends on the particular $\gamma_i$ and $\lambda.$ If we project the mirror symmetries onto the surface we find the effective representation $\hat{m}_{y,\text{eff}}=\sigma_1,$ which forbids any mass terms.

For path 2 one would expect that $k_z$ values between the two surface nodes lie in the quadrupolar phase since it is this part of the path that lies within the topological quadrupole region of the phase diagram. Hence, the bulk quadrupole moment is given as $q_{xy}=\frac{e}{2\pi}\mathcal{B}_z,$ where $2\mathcal{B}_z$ is the momentum space separation between the two surface Dirac nodes. For path 3 we see from Fig. \ref{fig:1}a that the portion of the path that lies in the topological quadrupole region is between pairs of nodes on opposite surfaces. Thus the quadrupole moment can be determined from the momentum-space differences between pairs of nodes on opposing surfaces. The resulting hinge states for these configurations are illustrated in Fig. \ref{fig:2}b.  We confirm these results numerically in Fig. \ref{fig:3}b,c where we calculate the surface polarization resolved over $k_z$. We also move path 2 (path 3) by tuning $\gamma_x \in [0.5, 1.5]$ ($[0.5, 1]$)while constraining $\gamma_y=1/4$ ($\gamma_y=-\gamma_x$) and calculate $q_{xy}.$ We compare it with the nodal separation formula above and find that they match.  

Finally, let us examine path 4. This path is topologically distinct from paths 2 and 3 because it encloses a bulk critical point; hence paths 2 and 3 cannot be smoothly deformed to path 4 without closing the bulk gap. We can parameterize this path using $\gamma_x=-1+1/2\cos k_z, \gamma_y=1+1/2\sin k_z.$ The TSM generated on this path is perhaps the most unusual because each surface normal to $\hat{x}$ or $\hat{y}$ harbors \emph{half} of a Dirac semimetal. This occurs because the path intersects the phase boundaries between the (1/2, 1/2) class and the (1/2, 0) and (0, 1/2) classes just a single time each. Hence, on each surface there will only be a single node in the energy spectrum, protected by mirror symmetry, and with a similar continuum Hamiltonian to $H_\text{surf-node}(k).$ However, because the path intersects the phase boundaries between (1/2, 0) and (0, 0), and (0, 1/2) and (0,0), there will be additional crossings in the Wannier bands not expressed in the energy spectrum. The structure of the Wannier band crossings is shown in Fig. \ref{fig:4}d,e, and interestingly this case shows spectral flow through the Wannier band values. To calculate the quadrupole moment of this type of TSM we consider the portion of the path that lies within the topological quadrupole phase, which in this case is between the nodal points located on two different surfaces. Thus, we expect this system to have a bulk quadrupole moment $q_{xy}=\frac{e}{2\pi}\mathcal{B}_z$ where $2\mathcal{B}_z$ is the momentum space separation between the a surface Dirac node on a surface normal to $\hat{x}$ and a surface Dirac node on a surface normal to $\hat{y}.$ In Fig. \ref{fig:1}b we have illustrated the hinge nodes that will appear, and how they are cutoff by the nodal positions on different surfaces. We confirm these results numerically in Fig. \ref{fig:3}d,h with similar plots as paths 2 and 3. In Fig. \ref{fig:3}h we shift path 4 by tuning $\gamma_x\in [-1.5, -0.5],$ while constraining $\gamma_y=-\gamma_x,$ and find that the calculated quadrupole moment matches the nodal separation formula. 

It is important to note that the three paths with surface TSMs are generated by the quadrupolar bulk topology, and are not equivalent to 2D semimetal layers glued onto otherwise insulating surfaces of a 3D gapped system. To illustrate this point let us consider the key feature of a bulk quadrupole moment which is that the magnitude of the hinge charge per unit length $q^\text{hinge},$ the surface polarizations $P^{\text{surf}}_x, P^{\text{surf}}_y$, and the quadrupole moment $q_{xy}$ satisfy the following relation\cite{raabbook,benalcazar2017A,benalcazar2017B}
\begin{eqnarray}\begin{aligned}
\label{eq:quad_formula}
q_{xy}=P^{surf}_x+P^{surf}_y-q^\text{hinge}.
\end{aligned}\end{eqnarray} To see that this is a bulk property that is insensitive to surface effects, let us imagine gluing a purely 2D Dirac semimetal to each of the two gapless surfaces of path 2, and tuning each 2D semimetal to have a pair of nodes separated in the $k_z$ direction, and aligned and degenerate with the existent nodes on those surfaces. Turning on tunneling between these layers and the surface will eliminate all of the Dirac nodes, and seemingly erase any phenomena associated with the surface semimetal. However, this is not the case. Adding the 2D Dirac semimetal on the surface will add to the surface polarization, and contribute charge to the hinges, proportional to the momentum space separation of its Dirac nodes\cite{ramamurthy2015}. As such, these two contributions to Eq. \ref{eq:quad_formula} exactly cancel, and the value for $q_{xy}$ is unmodified. Thus, despite such a drastic surface modification, the system still retains its bulk $q_{xy}.$ In a sense it remembers the locations of the ideal surface nodes because this information is contained the structure of the Wannier bands, which are derived from the bulk wavefunctions and independent of any surface termination.

So far we have evaluated the physical properties of these systems based on the locations of nodal points in the energy spectra to make contact with the extensive previous literature. However, from the argument above it is clear that one could tune the surface properties of paths 2 and 3 such that all of the surfaces are gapped, and yet there could still be a non-vanishing bulk quadrupole moment with the same magnitude.  Hence, in non-ideal cases where there have been modifications to the surface, and even for an ideal scenario with path 4, it may not be not obvious how to evaluate the bulk quadrupole moment using the conventional technique based on the momentum separation of energy nodes. In these cases it may be more natural to calculate the Wannier bands and use the Wannier nodal points to calculate the quadrupole moment. As a proof of concept, we performed this type of calculation for paths 2 and 4 in Fig. \ref{fig:4}, and the results match the calculations based on the nodes in the energy spectra from Fig. \ref{fig:3}. Specifically, we locate band crossings in the Wannier bands that occur at a value of $\nu=1/2,$ as it is precisely these crossings that are associated to band-crossings in the surface energy spectra for ideal surfaces. We shifted these paths by tuning the $\gamma_i$ exactly as in Fig. \ref{fig:3}, and we find that the quadrupole moment, and the momentum differences between Wannier nodal points match.

In conclusion, we have predicted several new types of topological semimetals that exhibit a bulk quadrupole moment. One type of semimetal has bulk nodes and gapped surfaces. A second type has a gapped bulk, but harbors a Dirac semimetal with an even number of nodes on one or more surfaces. The final type has a gapped bulk, but harbors half of a Dirac semimetal on multiple surfaces. Each of these give rise to mid-gap hinge states and hinge charge, as well as surface polarization. We also illustrated how the bulk quadrupole moments of these systems can be calculated from the momentum-locations of bulk or surface nodes in the energy spectrum. Finally, we showed that in some cases it is useful to examine nodes in the Wannier bands to extract the bulk quadrupole moment.

\acknowledgements{We thank W. A. Benalcazar, and B. A. Bernevig for discussions. ML thanks NSF  Emerging Frontiers in Research and Innovation NewLAW program Grant EFMA-1641084 for support. TLH thanks NSF CAREER Grant DMR-1351895 for support.}

\bibliography{quad_references}

\end{document}